 \documentclass[sigconf]{acmart}

\AtBeginDocument{%
  }

\setcopyright{rightsretained}
\copyrightyear{2026}
\acmYear{2026}
\acmDOI{}
\acmConference[UIST '26]{ACM Symposium on User Interface Software and Technology 2026}{November 2–5, 2026}{GM Renaissance Center in Detroit, MI}
\acmISBN{978-1-4503-XXXX-X/2018/06}

\usepackage{xspace}
\usepackage{cleveref}
\usepackage{algorithm}
\usepackage{algorithmic}
\newcommand{\systemname}{~\textbf{ArtAnno}\xspace}
\newcommand{\frameworkname}{BiHAA\xspace}

\usepackage{placeins}
\usepackage{float}
\usepackage{booktabs} 
\usepackage{tabularx}  

\usepackage{marvosym} 
\crefname{figure}{Fig.}{Figs.}
\Crefname{figure}{Fig.}{Figs.}
\begin{document}


\title[ArtAnno: Human-Agent Bidirectional Augmented Annotation in Artworks]{\textit{\systemname}: Annotating Implicit Semantics in Artworks through LLM Agent-Driven Bidirectional Human-AI Augmentation}

%

\author{Xiaoyan Gu}
\authornote{Both authors contributed equally to this research.}
\orcid{0009-0009-5379-985X}
\affiliation{%
  \institution{State Key Lab of CAD\&CG, Zhejiang University}
  \city{HangZhou}
  \state{Zhejiang}
  \country{China}
}
\email{xiaoyanGu@zju.edu.cn}

\author{Yifang Wang}
\authornotemark[1]
\orcid{0000-0001-6267-9440}
\affiliation{%
  \institution{Florida State University}
  \city{Tallahassee}
  \state{Florida}
  \country{USA}
}
\email{yifang.wang@fsu.edu}

\author{Wenqing Zheng}
\orcid{0009-0004-3293-5334}
\affiliation{%
  \institution{State Key Lab of CAD\&CG, Zhejiang University}
  \city{HangZhou}
  \state{Zhejiang}
  \country{China}
}
\email{wenqingzheng@zju.edu.cn}

\author{Haozhong Liu}
\orcid{0009-0008-8264-1683}
\affiliation{%
  \institution{State Key Lab of CAD\&CG, Zhejiang University}
  \city{HangZhou}
  \state{Zhejiang}
  \country{China}
}
\email{haozhong.24@intl.zju.edu.cn}

\author{Yixia Zheng}
\orcid{0009-0006-1943-1056}
\affiliation{%
  \institution{State Key Lab of CAD\&CG, Zhejiang University}
  \city{HangZhou}
  \state{Zhejiang}
  \country{China}
}
\email{yixia_zheng@zju.edu.cn}

\author{Peiyi Jiang}
\orcid{0009-0005-5799-9948}
\affiliation{%
  \institution{State Key Lab of CAD\&CG, Zhejiang University}
  \city{HangZhou}
  \state{Zhejiang}
  \country{China}
}
\email{12140023@zju.edu.cn}

\author{Wenjie Ning}
\orcid{0009-0008-4871-4586}
\affiliation{%
  \institution{School of Computer and Computing Science, Hangzhou City University}
  \city{Hangzhou, Zhejiang}
  \country{China}}
\email{32401194@stu.hzcu.edu.cn}

\author{Wei Zhang}
\authornote{Both authors are corresponding authors.}
\orcid{0000-0002-8321-4607}
\affiliation{%
 \institution{School of Computer and Computing Science, Hangzhou City University}
  \city{Hangzhou, Zhejiang}
  \country{China}}
\email{zw_yixian@hzcu.edu.cn}

\author{Wei Chen}
\authornotemark[2] 
\orcid{0000-0002-8365-4741}
\affiliation{%
  \institution{State Key Lab of CAD\&CG, Zhejiang University}
  \city{HangZhou}
  \state{Zhejiang}
  \country{China}}
\email{chenvis@zju.edu.cn}

\renewcommand{\shortauthors}{Trovato et al.}


\begin{abstract}
High-quality annotation of artworks is essential for computational art research, yet extracting implicit semantics remains challenging due to the reliance on culturally grounded meanings and deep contextual knowledge behind the images. 
Current AI-assisted annotation tools often lack assistance or rely on one-way workflows where experts have to perform extra manual calibrations to improve AI models, resulting in limited efficiency.
To address this, we propose \textit{Bidirectional Human-AI Augmentation (\frameworkname)}, a closed-loop framework in which skills and domain knowledge base evolve through real-time interaction and bidirectional HAI augmentation.
Informed by a formative study with 20 artwork annotators from different backgrounds, we implement this framework in \systemname, an artwork annotation system driven by a multi-agent architecture. 
The system includes a Proactive Agentic Support Module, where AI augments humans through semantic mining and label suggestion, and an Interaction-Driven Evolution Module, where human expertise continuously enhances the AI through distilling annotation trajectories into reusable experience. 
Evaluation through a user study and two case studies demonstrates that our framework and system improve annotation efficiency, enable knowledge accumulation, and reduce the effort of information seeking and verification for
annotators with limited domain expertise.
We conclude by discussing broader implications and future directions. 
\end{abstract}

\ccsdesc[500]{Human-centered computing~Interactive systems and tools}
\ccsdesc[500]{Computing methodologies~Multi-agent systems}
\ccsdesc[300]{Applied computing~Arts and humanities}

\keywords{Large Language Models, Multi-Agent Systems, Artworks, Annotation System, Human-AI Collaboration}

\maketitle

\section{Introduction}
High-quality annotation of artworks is fundamental for computational art research~\cite{DeArt, artbench}.
Beyond conventional annotation tasks such as object detection~\cite{lin2014coco}, artwork annotation requires analyzing the implicit semantics of the objects with culturally grounded meanings, symbolic associations, or interpretive implications invisible to pixels alone~\cite{Dubourg2024CulturalAnnotation, artEmis, artbench}. 
For instance, in \textit{The Last Supper}, conventional annotation may label the salt cellar simply as a physical object, overlooking its cultural symbolism of trust and purity in Renaissance contexts.
This results in a failure to uncover the spilled salt as a visual metaphor for broken fellowship and Judas’s impending betrayal. 
Bridging this gap remains the ``last mile'' of artwork analysis, requiring tools that can integrate and record visual evidence with deep contextual knowledge~\cite{visinfo4, visinfo6}.

Extracting implicit semantics remains hindered by low efficiency and high entry barriers in artwork annotation tasks. 
First, most annotation tools do not provide assistance during the artwork annotation process, which requires considerable effort to identify the semantic meanings of depicted objects~\cite{CVAT2024, LabelStudio2026}.
While a few annotation tools enhance efficiency through interactive algorithmic assistance (e.g., human pose estimation~\cite{semianno}), they offer little support for the semantic discovery of artworks.
Second, the inconvenience of experts to leverage prior knowledge leads to repetitive tasks, which diminishes efficiency.
Although researches attempt to lower the entry barrier and enhance efficiency by capturing domain expertise through interactive parameter adjustment (e.g., manual configuration of feature importance~\cite{markup}), such mechanisms often introduce extraneous interaction costs that inadvertently increase the workload, yielding limited improvements in overall efficiency.

Recent research in Human-AI collaboration (HAI), empowered by rapid
advances in Large Language Models (LLMs), has positioned agents as
assistants or co-creators; while these agents iteratively refine
their abilities through skill curation and reflection on execution
traces~\cite{evoAgents, coevolution2}.
However, these paradigms remain unidirectional: either AI assists
humans in tasks, or humans facilitate AI evolution through extra
manual calibration.
The decoupling of these stages leads to a fragmented workflow, consequently requiring additional manual effort to support model improvement.
To address this question, building on existing literature on AI agents and HAI~\cite{postermate, coevolution} and insights from internal experts, we first propose a \textit{Bidirectional Human-AI Augmentation (\frameworkname)} framework, which conceptualizes Human-AI collaboration as a bidirectional augmentation process rather than a one-way mechanism.
Within this framework, AI provides proactive support for task completion, while human interactions are captured and transformed into reusable expertise to continuously refine the system. 
This bidirectional synergy creates a closed loop where humans and AI capabilities mutually augment each other across interaction rounds.

To validate our framework and identify design requirements, we conducted a formative study with 20 artwork annotators from diverse backgrounds using a Wizard-of-Oz~\cite{Kelley1984WizardOfOz} methodology.
The results validate the framework in real-world scenarios and reveal specific user requirements.
Results show that users require proactive recommendations at the initial stage to reduce exploration overhead, followed by interactive multimodal explanations for verification, and expect the system to internalize both domain knowledge (e.g., cultural symbolism) and procedural strategies for continuous AI refinement.
Based on these findings, we formulate four system requirements.
Guided by the \textit{\frameworkname} framework and requirements, we designed and implemented \systemname, a system tailored for artwork annotation.
The system consists of two multi-agent-driven modules: the Proactive Agentic Support Module, which supports semantic mining and proactive label suggestion, with multimodal justifications, and the Interaction-Driven Evolution Module, which analyzes annotation results and user interaction trajectories, internalizing domain knowledge into a Knowledge Base and procedural expertise into a reusable Skill Library.
This continuous loop enables effective assistance and continuous knowledge accumulation, improving system performance for future tasks.
Finally, we validated \systemname\ through two case studies and a user study, showing that it improves annotation efficiency, enables knowledge accumulation and reuse, supports cross-domain transfer.
In summary, this work makes three main contributions:
\begin{itemize}
    \item \textbf{The \textit{\frameworkname } Framework:} We propose a novel conceptual framework for \textit{Bidirectional Human-AI Assistance}, establishing a reciprocal loop that enables real-time intelligent support and the continuous refinement of expert expertise.
    
   \item \textbf{The \systemname System:} We develop \systemname, a multi-agent driven system that operationalizes the \textit{\frameworkname} framework in the context of artwork annotation.
    
    \item \textbf{Empirical Validation:} We conduct comprehensive evaluations involving user study and case study. Our results demonstrate the system's effectiveness in enhancing annotation efficiency.
\end{itemize}

\section{Related Work}
In this section, we review two areas relevant to our research: annotation systems and the Human-LLM Agent collaboration.

\subsection{Annotation System}
Artworks refer to visually presented imagery that conveys cultural meanings or deeper aesthetic expressions, ranging from traditional paintings to poster and comics~\cite{berger1972ways, mitchell1994picture}; due to art expression, artwork annotation requires analyzing the implicit semantics of objects with culturally grounded meanings, symbolic associations, or interpretive implications that are invisible at pixel level~\cite{Dubourg2024CulturalAnnotation, artimgannoframework}.
Annotating such information is challenging, as it is both time-intensive and requires annotators with domain expertise in art.

In recent years, a number of annotation tools have been developed to improve annotation efficiency~\cite{DeepEdit, sam2, PAMSNet, SemanticAnno}.
Among them, general image annotation systems (e.g. VIA~\cite{VIA_tool}, Label Studio~\cite{LabelStudio2026}) primarily improve efficiency through streamlined interfaces, configurable annotation templates, and shortcut-based interactions, whereas workflow-optimized systems (e.g. CVAT~\cite{CVAT2024}) further accelerate annotation through interpolation, task management, and semi-automatic annotation support~\cite{vista, sam3, sam2}.
Several data visualization approaches present visual clusters and shared characteristics of samples, enabling pattern discovery and annotation propagation for batch annotation~\cite{vista, AnnoLens, CalliVA,KALE}.

These methods improve general annotation efficiency, yet they are insufficient for domain-oriented annotation tasks due to the lack of domain knowledge. 
To address this issue, a number of domain-specific annotation systems have been developed to improve annotation efficiency by incorporating domain knowledge into the annotation process~\cite{medSAM, llava, medbiaser, AnnoLens}. For example, MedSAM2~\cite{medsam2}, a medical annotation method that embeds medical knowledge into segmentation models to support prompt-based annotation of 3D medical images.
In the humanities domain, CataAnno~\cite{cataanno} integrates historical knowledge by visualizing relationships between historical entries, thus assisting annotation and reducing the need for specialized knowledge. 
In the art domain, systems such as Visual Narratives~\cite{visNarra} and ArtSeek~\cite{artseek} incorporate art historical knowledge by focusing on classification based on styles or periods~\cite{artimgannoframework, eCul, SemanticAnno, visinfo3}.
However, the domain knowledge embedded in these systems is usually fixed and limited, and often requires additional training or model updates to maintain, making it difficult to continuously acquire expert knowledge and effectively reduce the reliance on domain experts.

With the emergence of LLM-based agents, recent studies have explored agent-based approaches to assist annotation and improve efficiency~\cite{humanLLMcoannotation, IAI, ModelingCollaborator}. 
For example, CrowdAgent~\cite{crowdagent} employs LLM-based agents to decompose annotation tasks, provide interactive guidance, and automate parts of the annotation workflow, thereby reducing manual effort. 
Meanwhile, other approaches employ active learning or Human–AI feedback loops to iteratively improve annotation performance~\cite{medbiaser}. For instance, KMTLabeler~\cite{KMTLabeler} uses active learning to select informative samples and incorporates human feedback to update the model during the annotation process, thereby improving annotation efficiency and model performance. However, these approaches are often designed as separate workflows or require additional user intervention, introducing interaction overhead that can increase the workload of the annotators and lead to limited overall efficiency gains.

Generally, current annotation tools struggle with implicit semantics and require manual effort and expertise. Our work addresses these issues through a Human–Agent collaborative framework.
By enabling a bidirectional interaction loop, our system enhances efficiency and allows for iterative, within-session evolution.

\subsection{Human–LLM Agent Collaboration}

Recent works have investigated the role of LLM agent tools in assisting humans and boosting productivity in diverse settings, including creative design~\cite{postermate, creativityAgent}, software development~\cite{softwareAgent}, interactive decision-making~\cite{decisionAgent, visinfo1}, and scientific discovery~\cite{shao2026sciscigpt, wang2026figures, shi2026survey}.
AI agents like OpenClaw~\cite{OpenClaw2026} and VisionGPT~\cite{Kelly2024VisionGPT} have become increasingly tailored and multimodal for analyzing data and deriving insights~\cite{multiagentmultimodal}. 
With human guidance, PosterMate can use audience-driven persona agents to assist poster design through customized discussions~\cite{postermate}; DuoDrama interacts with screenwriters through perspective-aware feedback to help scriptwriting~\cite{DuoDrama}. 
These advancements demonstrate the capacity of agents to actively augment human decision-making through domain-specific insights. 
    
In turn, humans can also help agents evolve~\cite{coevolution, coevolution2, coevolution3}. 
Agents can summarize operation-level user interactions into task-level skills, enabling real-time self-improvement without relying on offline parameter updates~\cite{surveyselfevolvingagentswhat, evoAgents}. According to Xiang et al., the Environment‑Centric Self‑Evolution supports refined skill growth and experience-driven continual learning~\cite{evoAgents}. Recent studies, such as XSkill~\cite{xskill} and AutoSkill~\cite{autoskill}, further provide tools for this lifelong learning via self-evolution of reusable skills~\cite{coevolution, visinfo2}. 
The evolution process highlights a dynamic in which humans also actively contribute to the continual optimization of AI agents.

Existing Human--AI collaboration paradigms improve efficiency and model
performance, but the two directions remain largely decoupled: either AI
assists humans, or humans provide feedback to improve AI in separate
stages, resulting in fragmented workflows and extra manual effort for
model updates.
Even recent skill-accumulating agents such as XSkill~\cite{xskill},
AutoSkill~\cite{autoskill}, and environment-centric
self-evolution~\cite{evoAgents} update skills only in batched phases or
upon explicit triggers, reuse workflows rather than domain knowledge,
and are evaluated mainly on automated metrics.
In this study, we propose a Human--AI collaborative framework that forms
a reciprocal loop between agent assistance and human knowledge
contribution. In contrast, our framework updates skills in real
time within the same workflow, reuses both workflows and domain
knowledge through a hierarchical, structure-aware Skill Library, and is
validated through human-centered user studies.
A more structured table comparing our method with existing method is provided in the supplementary material.

\section{Study Design}
In this section, we first propose a \textit{Bidirectional Human–AI Augmentation Framework (BiHAA)} informed by the limitations identified in current HAI literature and theories, as well as real-world insights gathered from our internal domain experts in artwork annotation.
To validate the rationale of our framework and to understand the specific requirements of artwork annotation, we conducted a formative study with 20 annotators from diverse backgrounds.
Based on the study findings, we validate the framework and derive four design requirements to inform the system design.

\subsection{\textit{BiHAA}: Bidirectional Human-AI Augmentation Framework}

\begin{figure}[!t]
    \centering
    \includegraphics[width=1\linewidth]{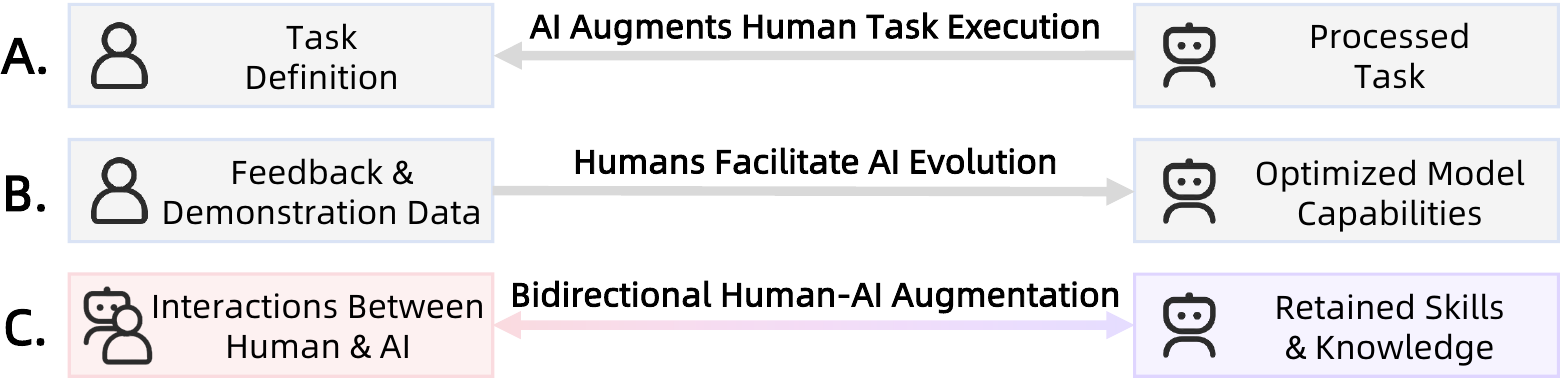}
    \caption{Comparison between the \textit{Bidirectional Human-AI Augmentation Framework (BiHAA)} and traditional frameworks. (A-B) Left block represents input, right one represents output. (C) Both blocks serve as input and output.}
    \label{fig:framework}
    \vspace{-3mm}
\end{figure}

In recent years, AI agents have evolved from passive command executors into collaborative partners that continuously improve through human feedback, along with skill management, tool use, and reflection on past tasks to accumulate reusable knowledge~\cite{coevolution, visinfo5}. 
However, existing research primarily follows two independent paths: one focuses on how AI assists humans during tasks~(\cref{fig:framework}(A)), while the other examines how humans improve AI capabilities through feedback, correction, or demonstration~(\cref{fig:framework}(B)). 
As a result, these processes are often separated, and human knowledge generated during task execution is not effectively captured and reused for future tasks.

Feedback from our internal expert annotators further highlights this issue: due to limited pre-annotation accuracy, annotators often need to consult external references and repeatedly correct similar errors, while the model cannot learn from these corrections in real time. 
This suggests that users expect AI not only to assist with the current task, but also to learn from human feedback to improve future system performance.
While a few recent studies have started to facilitate bidirectional Human-AI communication to align agent responses with human intent, they often stop short of mutual enhancement. 
We ask whether this bidirectional loop can be extended beyond intent communication to enable bidirectional augmentation: 
can AI assist humans in the current task while simultaneously internalizing human expertise to accelerate future tasks?

Here, we propose the \textit{Bidirectional Human–AI Augmentation Framework (BiHAA)}, which conceptualizes Human-AI collaboration not as a one-way assistance mechanism, but as a bidirectional augmentation process where the system continuously evolves by accumulating human expertise through ongoing usage (\cref{fig:framework}(C)). 
Users first complete predefined tasks with AI assistance, where the AI functions as either a supportive assistant or a collaborative partner to augment users' capabilities.
During task execution, the system records user actions and Human-AI interaction processes as logs and further extracts expert heuristics from these logs, transforming them into reusable skills~\cite{xskill} and knowledge that are integrated into a shared skill or knowledge base. 
In subsequent similar tasks, the system can query and invoke relevant existing expertise to provide informed support, enabling experts to complete tasks more efficiently. 
Through this mechanism, the system transforms one-off Human-AI interactions into a perpetual cycle of continuous knowledge accumulation and collaborative optimization.

\subsection{Formative Study}
To validate the \textit{BiHAA} framework and understand how to apply our framework to domain-specific challenges in artwork annotation, we conducted a two-stage formative study.
Participants first completed a cultural semantic annotation task for traditional Chinese paintings, followed by a semi-structured interview.
The interviews were conducted with two primary objectives: (1) to examine user requirements for agent assistance from a 3W1H (Who, What, When, How) perspective based on the Kipling Method~\cite{Kipling1902JustSo}; and (2) to identify in the artwork annotation scenario, which experiences and decision-making processes embedded in user workflows should be captured and accumulated to enable the continuous enhancement of system capabilities for future tasks.
The following sections detail the study's Participants and Procedure as well as the Study and Results. 
This study was conducted with the approval of our institutional ethics committee.

\subsubsection{Participants and Procedure.}
We recruited 20 annotators from diverse domains through open
recruitment. They were experienced in annotation but did not
necessarily have an art background, matching our target users:
skilled annotators without formal art training. Additional
participant details are in the supplementary materials.
This formative study had two phases. In the first, participants
identified cultural semantics in Traditional Chinese Painting using a
prototype in a Wizard-of-Oz setup~\cite{Kelley1984WizardOfOz}, where
a human operator simulated the bidirectional evolution process by
collecting knowledge, updating the system between rounds, and
applying it in subsequent rounds. For example, participants
interpreted ``longevity'' from the co-occurrence of a cat and a
butterfly, due to their homophonic association with ``maodie'' in
Chinese; the operator then distilled this behavior into an
annotation rule and added it to the assistance panel for later
rounds. The operator intervened only between rounds, while in-round
assistance followed the pre-prepared panel. This maps directly onto
the BiHAA framework that ArtAnno later automates: the panel
corresponds to Proactive Agentic Support, and the operator's manual
knowledge extraction to Interaction-Driven Evolution (see Appendix
for the full protocol).
Each participant annotated twenty works, during which we recorded
both total time and the specific behavioral processes.

\subsubsection{Study Results.}
This section presents the empirical observations and key findings derived from the study, providing a detailed analysis of user requirements for AI assistance.

\textbf{Validation of the Framework:}
During the experiment, we observed that as participants continued using the system, the Wizard-of-Oz operator simulated system updates by incorporating knowledge collected from users’ previous searches and interactions.
Reusing this accumulated knowledge significantly reduced the time users spent on repeated information retrieval.
In the interviews, participants also noted that the system could learn from their interactions without requiring additional effort and provide relevant knowledge in later stages, which improved their efficiency in subsequent tasks.
These observations support the effectiveness of our framework: while users complete their tasks, the system evolves by learning from user behavior and knowledge, which can improve task efficiency and provide cumulative benefits in repeated or similar tasks within a
session.
At the same time, some participants suggested that the system should not only extract knowledge but also learn their search strategies and operational workflows.

\textbf{User Feedback:}
Based on the formative study, we summarize user's feedback from two aspects: 
(1) users’ needs for AI assistance during the task, and 
(2) the types of knowledge that should be captured and accumulated from the annotation process.

\textit{Feedback 1: User Needs for AI Assistance (\textbf{3W1H}).}
From the interviews, participants expressed expectations regarding the role and functionality of AI assistance during the annotation process. 
First, regarding the system’s role (\textbf{Who}), most participants viewed AI not merely as a tool, but as a collaborative partner capable of supporting decision-making and offering suggestions during the task.
Second, concerning the timing of intervention  (\textbf{When}), participants preferred AI assistance to be available at the beginning of the task for initial guidance, and then on-demand throughout the process whenever they encountered difficulties.
Third, about the output content  (\textbf{What}), participants expected AI to provide label recommendations along with explanations or reasoning, rather than offering labels without context.
Finally, regarding the interaction modality  (\textbf{How}), participants preferred assistance that both captured their attention for annotation tasks, such as pop-up suggestions, and provided structured, accessible information, like text and reference materials anchored in a system sidebar, to help them verify and understand the suggested annotations.

\textit{Feedback 2: Knowledge and Process Worth Capturing.}
In addition to AI assistance needs, participants also highlighted that certain types of domain knowledge and task processes should be captured and accumulated during the annotation process. 
Specifically, the study highlights the importance of domain knowledge, including symbolic meanings, cultural context, and the logic of interpretation.
Moreover, participants emphasized that the annotation process itself, including search strategies, decision-making steps, and verification methods, reflects expert experience and should be recorded and modeled. 
Capturing both domain knowledge and task processes can support future annotations and help less experienced annotators perform tasks more efficiently.

\subsection{System Requirements}

Based on the results above, we derive the following design requirements to guide the system design.

\textbf{R1: Recommend clusters of images with shared annotation characteristics.}
Annotators frequently encounter images with similar characteristics, yet these images appear in a disorganized order, imposing a high cognitive burden. 
The system should cluster images with similar visual and semantic characteristics and provide shared label recommendations, enabling batch annotation and reducing redundant effort across similar instances.

\textbf{R2: Proactive Label Recommendation and On-Demand Knowledge Query.}
The system should provide proactive suggestions at appropriate stages, such as at the beginning of annotation or when users encounter difficulties, presented as attention-guiding pop-up prompts.

\textbf{R3: Support Verification of Recommended Labels through Multimodal Evidence.}
Since annotation decisions often require validation from multiple sources, the system should provide multimodal information, such as text descriptions, visual references, and external sources, to help users verify AI suggestions.

\textbf{R4: Distill and Reuse Knowledge and Exploration Trajectories.}
The system should capture expert knowledge as well as users’ interaction trajectories and decision processes, and transform them into reusable skills or knowledge to support future tasks.

\begin{figure*}[t]
  \includegraphics[width=1\textwidth]{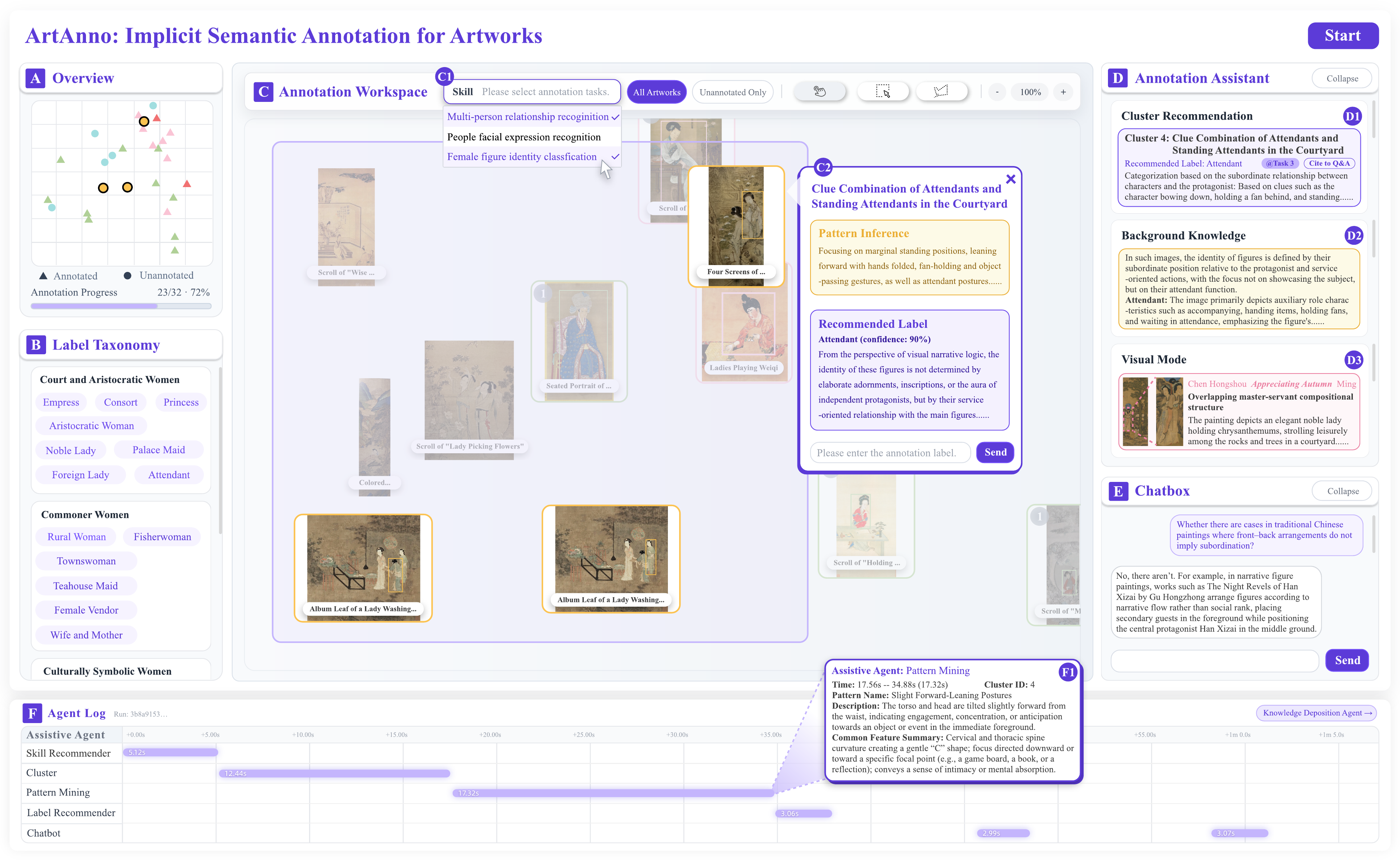}
  \caption{ArtAnno System Interface for Annotation of Female Identity and Experience Accumulation in Traditional Chinese Paintings. Exploration begins with (A) Overview, which displays the clustering results of artworks based on the semantic hierarchy in (B) Label Taxonomy. Selected clusters lead to (C) Annotation Workspace, where users can perform detailed annotation, and (D) Annotation Assistant proactively provides cluster recommendations, background knowledge, and visual mode to aid the process. Throughout the workflow, (E) Chatbox enables Human-Agent dialogue and (F) Agent Log exposes agent behaviors in two stages, which jointly help users verify and understand the annotation results.}
  \label{fig:teaser}
\end{figure*}

\section{ArtAnno}
This section presents the design of \systemname.
We first provide a system overview, followed by a detailed description of the data preparation and the multi-agent architecture that drives the system workflow.
Finally, we introduce the frontend interface.

\subsection{System Overview}
The system is powered by a multi-agent collaborative backend designed to support annotation and knowledge construction workflows. 
Based on the proposed framework, the backend is divided into two core components: the Proactive Agentic Support Module and the Interaction-Driven Evolution Module. 
These two modules interact continuously to support user interaction and annotation processes.
The frontend interface, implemented using React.js, consists of two primary areas: a User Annotation Workspace and an Agent Log Display Panel, which provide real-time feedback and transparency for agent behavior. 
The backend is implemented with Flask, while the LLM driving the agents is GPT-5, orchestrated through the LangChain agent development framework.

\subsection{Data Preparation}

The data preparation stage aims to construct both visual and semantic representations of artworks to support subsequent clustering and analytical tasks. 
First, an object detection algorithm, for which we use YOLO-World-V2~\cite{cheng2024yolo}, is applied to the artworks to generate bounding boxes and corresponding basic labels for visual elements within the paintings.
By leveraging the original image, associated metadata, and cropped BBox regions, the LLM generates structured textual descriptions across two hierarchical levels: (1) artwork-level descriptions that capture the overall content and context of the painting to convey background information for better identity inference, and (2) unit-level descriptions that describe the semantic meaning, visual attributes, and contextual role of each annotated unit. 
The resulting multimodal data consists of bounding boxes, basic labels, and language descriptions, which serves as the foundation for downstream processes.

\subsection{Backbone Multi-Agent System}
The system backend adopts a multi-agent collaborative architecture comprising two core modules: the Proactive Agentic Support Module (\cref{fig:workflow}(A)), which provides intelligent support to enhance annotation efficiency; and the Interaction-Driven Evolution Module (\cref{fig:workflow}(B)), which captures and transforms user interaction trajectories into reusable knowledge to drive the system's evolution.
Detailed agent designs for these modules are described below.

\subsubsection{Proactive Agentic Support Module}

The Proactive Agentic Support Module enhances annotation efficiency and reduces cognitive burden by providing proactive support during annotation. It consists of four coordinated agents: a Skill Recommendation Agent, a Cluster and Pattern Mining Agent, a Label Recommendation Agent, and a Chatbot Agent.

\textit{Multi-Agent Workflow.}
First, the task description and the descriptions of each skill are provided to the Skill Recommendation Agent (\cref{fig:workflow}(1)), which selects the top three relevant skills based on the task requirements and recommends them to the user. 
After the user selects the required skills, the selected skills are assigned to the Cluster and Pattern Mining Agent (\cref{fig:workflow}(2)) and the Label Recommendation Agent (\cref{fig:workflow}(3)) to guide the subsequent pattern mining and label recommendation processes, ensuring that the recommendations align with the task context and annotation goals.
Next, the preprocessed data are sent to the Cluster and Pattern Mining Agent for clustering and pattern mining.
During clustering, the agent can call text similarity embedding models and image similarity embedding models as tools to compute similarity and group images into clusters accordingly. 
Based on the initial clustering results, the agent further refines the clusters into smaller sub-clusters and summarizes the shared characteristics of each cluster. 
The clustered images and their pattern descriptions are then passed to the Label Recommendation Agent (\cref{fig:workflow}(3)), which matches image features, cluster-level patterns, and label semantics to generate recommended labels along with corresponding reasoning.
The recommended labels, reasoning process, and conclusions are then presented on the front-end interface for users to reference during annotation. 
During the annotation process, users can also consult the Chatbot Agent (\cref{fig:workflow}(4)) at any time by asking questions about a single image or an image cluster to obtain explanations and suggestions, label meanings, or annotation decisions.

\textit{Label Recommendation and Prioritization.}
To leverage shared patterns among similar samples and improve recommendation reliability, we perform label recommendation at the cluster level by matching cluster patterns with label definitions, rather than treating it as an instance-level classification task.
For each cluster, the system first summarizes cluster-level patterns, including object co-occurrence, spatial relationships, inferred semantic themes, and patterns relevant to the current task.
These pattern summaries are then aligned with label definitions in the label library to compute a matching score.
The matching score considers the semantic consistency between the cluster pattern and the label definition, as well as prior support from the knowledge base.
Labels are ranked according to this matching score, and the top-ranked labels are recommended for each cluster along with corresponding rationales.

In addition, we introduce a cluster prioritization mechanism to determine which clusters should be presented to users first for verification.
Since users cannot review all clusters simultaneously, each cluster is assigned a priority score based on label uncertainty, task relevance, and potential impact.
Label uncertainty measures how ambiguous the label ranking results are, where similar scores among candidate labels indicate higher uncertainty.
Task relevance measures how closely a cluster is related to the current task. 
Potential impact measures how many samples may be affected if the cluster is corrected, reflecting the influence of user feedback on subsequent annotation.
The cluster priority score is defined as follows:
\begin{equation}
Priority(c) = \lambda_1 S_{\text{uncertainty}}(c) + \lambda_2 S_{\text{task}}(c) + \lambda_3 S_{\text{impact}}(c)
\end{equation}
where $S_{\text{uncertainty}}$ denotes Label uncertainty, $S_{\text{task}}$ denotes
task relevance, $S_{\text{impact}}$ denotes potential impact, and
$\lambda_1, \lambda_2, \lambda_3$ are weighting parameters.
To set the weights, we compared several configurations---
uncertainty-heavy $(0.6,0.2,0.2)$, relevance-heavy $(0.2,0.6,0.2)$,
and impact-heavy $(0.2,0.2,0.6)$. The top-3 labels remained over 75\%
consistent across settings, showing that recommendations are robust
rather than fragile; the remaining differences lie in which clusters
are surfaced first, confirming that the weights meaningfully shape
prioritization. We adopt $\lambda_1{=}0.4$, $\lambda_2{=}0.3$,
$\lambda_3{=}0.3$, which prioritizes label uncertainty so that the
most ambiguous clusters are verified first while keeping the three
factors balanced.
Clusters are ranked according to the priority score, and the system proactively
presents the highest-priority clusters for user inspection, transforming label
recommendation into an active decision support process.

\subsubsection{Interaction-Driven Evolution Module}

The Interaction-Driven Evolution Module is designed to continuously improve the system by capturing implicit knowledge embedded in users` interaction behaviors and feedback, and transforming it into reusable annotation skills and accumulated knowledge.

\begin{figure*}
    \centering
    \includegraphics[width=1\textwidth]{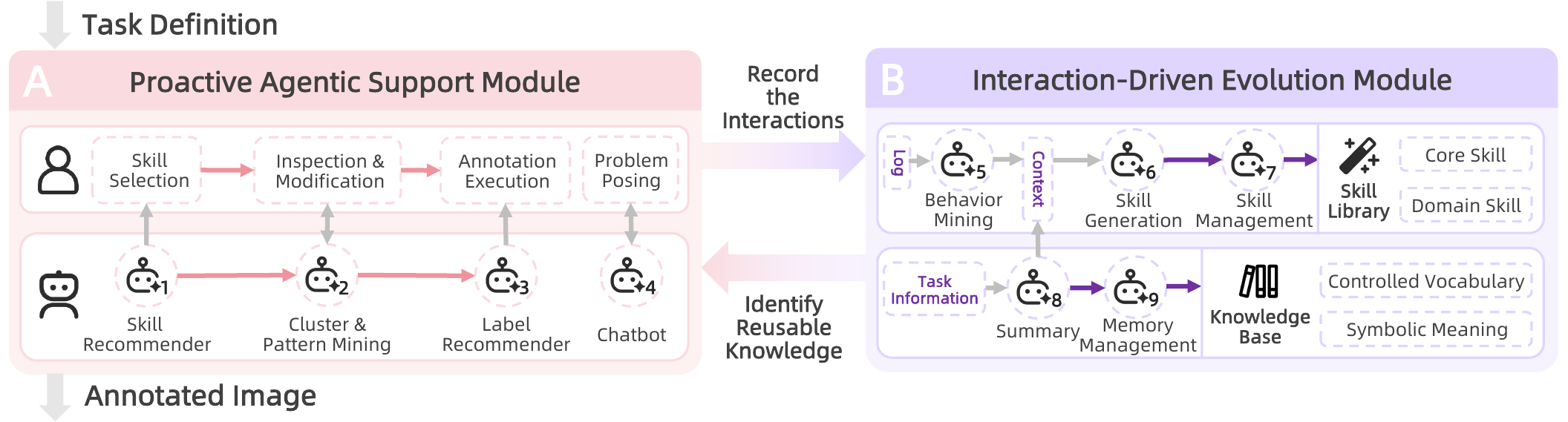}
    \caption{Overview of the ArtAnno System. The system integrates two core components: (A) Proactive Agentic Support, which optimizes annotation efficiency through a multi-agent pipeline for skill recommendation, data clustering, and label recommendation; and (B) Interaction-Driven Evolution, which transforms user trajectories into reusable assets via behavior mining, autonomous skill generation and management, and persistent knowledge base updates.}
    \label{fig:workflow}
\end{figure*}

\textit{Workflow:} 
We define a \emph{skill} as a structured, reusable procedure that encodes an
annotation workflow, distilled from human--AI interactions and stored in the Skill
Library for later reuse.
Initially, the Behavior Mining Agent (\cref{fig:workflow}(5))
performs behavior mining to analyze user actions and feedback, such as label
acceptance, correction, submission, and reasoning traces, extracting useful data
that reflects the users' habits and task requirements.
Based on this mined data, the Skill Generation Agent (\cref{fig:workflow}(6)) generates new skills, which are then stored in the Skill Library for future use and management. These skills are continuously refined by the Skill Management Agent (\cref{fig:workflow}(7)) to ensure they adapt to evolving user needs and tasks.
Additionally, the Summary Agent (\cref{fig:workflow}(8)) summarizes task information and updates the memory management process, while the Memory Management Agent (\cref{fig:workflow}(9)) incorporates the newly acquired knowledge into the Knowledge Base. 
This process allows the Interaction-Driven Evolution Module to evolve dynamically, not relying solely on predefined rules or expert input, but instead leveraging real-world annotation practices and continuous human feedback to enhance its capabilities.

\textit{Skill Management with Structure-Aware Merging.}
As the number of skills grows with continuous task execution, storing all generated skills can lead to redundancy and inefficiency, especially when skills share similar structures but differ in specific components.
To address this issue, we introduce a structure-aware incremental skill merging mechanism to assist the Skill Management Agent.

Each skill is represented as a structured document composed of multiple modules.
Specifically, an Overview describes the purpose and application scenarios of the skill, while Preconditions specify the required environments and dependencies.
The Workflow defines the step-by-step process of input, processing, and output.
In addition, Best Practices capture practical guidelines and experience, Examples provide representative usage cases, and Troubleshooting covers common issues and their corresponding solutions. 
Based on this modular representation, the Skill Management Agent performs structure alignment between a newly generated skill and existing candidate skills. 
It then conducts module-level difference analysis to identify shared components and divergent modules.
When two skills are largely consistent, their shared structure is preserved, while differences in specific modules are treated as conflicts or variations.
These conflicts are resolved through merging, replacement, or version branching, resulting in an updated skill with a new version.
The detailed algorithm for this process is outlined in Algorithm \ref{alg:skill_merge}.
This design enables skills to evolve incrementally through structured updates rather than redundant accumulation.

\begin{algorithm}
\caption{Structure-Aware Skill Merging}
\label{alg:skill_merge}
\small
\begin{algorithmic}[1]
\REQUIRE New skill $S_{new}$, skill library $\mathcal{S}$, threshold $\theta$
\ENSURE Updated skill library $\mathcal{S}$

\STATE Retrieve candidate skills $\mathcal{S}_c$ similar to $S_{new}$

\FOR{each skill $S_i \in \mathcal{S}_c$}
    \STATE Align module structures of $S_{new}$ and $S_i$
    \STATE $\Delta \leftarrow \textsc{ModuleDiff}(S_{new}, S_i)$
    
    \IF{$\Delta < \theta$}
        \STATE Identify conflicting modules in $\Delta$
        \STATE Resolve conflicts via merging, replacement, or branching
        \STATE $S_{merged} \leftarrow \textsc{Merge}(S_{new}, S_i, \Delta)$
        \STATE Update version of $S_{merged}$
        \STATE Replace $S_i$ with $S_{merged}$ in $\mathcal{S}$
        \RETURN $\mathcal{S}$
    \ENDIF
\ENDFOR

\STATE Add $S_{new}$ as a new skill into $\mathcal{S}$
\RETURN $\mathcal{S}$
\end{algorithmic}
\end{algorithm}

\subsection{Frontend Interface}

The system interface consists of six main components.
The Overview presents a layout of the artworks arranged according to their similarity in the task-specific label space. Different colors indicate clusters, circles represent unannotated images, triangles denote annotated ones, and the overall annotation progress is displayed (\cref{fig:teaser}(A)). 
The label taxonomy stores the labeling schema  (\cref{fig:teaser}(B)). 
The central annotation workspace serves as the primary annotation area, where images are visualized through clustering and dimensionality reduction; recommended clusters are highlighted with distinct backgrounds, and pop-up panels provide suggested annotations  (\cref{fig:teaser}(C)).
On the right, the assistance panel (\cref{fig:teaser}(D)) supports annotation from three perspectives, including cluster-based recommendations (\cref{fig:teaser}(D1)), background context (\cref{fig:teaser}(D2)), and visual patterns (\cref{fig:teaser}(D3)). 
At the bottom, a chatbox allows users to ask questions by selecting either a cluster or an individual image (\cref{fig:teaser}(E)). 
Additionally, an agent log records the agent’s reasoning process, enabling users to monitor, intervene, and provide feedback in real time to prevent error propagation (\cref{fig:teaser}(F)).

\section{Evaluation}
We evaluate our system through a user study and two case studies, focusing on the effectiveness of Agent in assisting human creative workflows and the capacity for human interventions to drive iterative Agent refinement.

\subsection{User Study}
Our user study consisted of two parts: a controlled experiment and a semi-structured interview.

\textbf{Apparatus}:
The ArtAnno system used in the study is driven by GPT-5 accessed through OpenRouter
with temperature set to 0 for deterministic behavior. Processing a set of 30 images
requires approximately 0.95M input and 110k output tokens in total. Image
preprocessing takes about 10s per image and is performed ahead of time, so latency
during interactive annotation is negligible.

\textbf{Participants:} We recruited 12 volunteers for the user experiment. 
All participants had prior experience with annotation tools, with a mean self-reported proficiency of 3.67 on a 5-point scale (1–5). 
Participants’ level of familiarity with artworks varied, yielding an average self-assessment score of 2.33. 
Each participant received \$15 as compensation upon completing the experiment.

\textbf{Control Group:} 
This study includes three experimental conditions:
(1) Condition 1 Baseline (C1). The system provides users with images, metadata, a predefined labeling taxonomy, and pre-annotated bounding boxes, while allowing the use of external search engines or AI tools for assistance (see Supplementary Materials for the interface illustration).
(2) Condition 2 Ablated System (C2). This configuration utilizes the core system without the dynamic evolution component. Participants are restricted to the system’s native feature set and operate without external information retrieval or third-party AI assistance, ensuring task is completed using only the provided local tools.
(3) Condition 3 Full System (C3). This condition uses our complete system, integrating the Interaction-Driven Evolution Module to provide knowledge-based assistance throughout the annotation process, while leveraging evolved skills to streamline the workflow and improve output quality.
C1 simulates the current real-world baseline, where annotators
rely on external search engines or AI tools. In C2 and C3, the
system's agent natively handles chat and web search, so external
tools are disabled to prevent them from confounding the comparison
and to isolate the contribution of our modules.

\textbf{Data and Task: }In the data and task design, each condition required annotating 30 female figures in Traditional Chinese Painting.
Reference labels were constructed by two domain experts (each with over five
years of experience in art history), who independently labeled all samples and
resolved disagreements through discussion to form the final reference set.
The images were divided into three groups of comparable difficulty, each assigned to one condition. Participants completed all three conditions, with the order counterbalanced to mitigate order effects; since each condition used a distinct image
group, label-specific transfer is limited and any residual learning
effect is evenly distributed across conditions, ensuring a fair
comparison between conditions.
Annotation time for each condition was recorded for analysis. To ensure consistency, all participants followed the same annotation guidelines, labeling taxonomy, and task instructions, and completed the tasks independently.
After completing each condition, participants were asked to fill out the System Usability Scale (SUS)~\cite{Brooke1996SUS} and a set of task-specific questions using a 5-point Likert scale. This was followed by a 30-minute semi-structured interview to collect qualitative feedback.

 \textbf{Results and Analysis: }
 We summarize the results and analysis from five aspects: system usability, task-specific evaluation, efficiency, system evolution, and insights from interviews. Values are reported as mean ($M$) and standard deviation ($SD$).

\textit{System Usability.}
We evaluated the system’s usability using the System Usability Scale (SUS)\cite{Brooke1996SUS}, which utilizes a 5-point Likert scale. The overall SUS score was 84.17, indicating excellent usability.
The detailed SUS scores are provided in the supplementary materials.
Overall, users found the system intuitive and easy to use, as the mean score for most dimensions exceeded 4.0. 
Specifically, participants rated ease of use (M = 4.08, SD = 0.65) and confidence in using the system (M = 4.08, SD = 0.65) highly, reflecting a positive user experience.
A few reverse-coded items related to system complexity and consistency received lower ratings, with scores of 1.58 (SD = 0.79) and 1.08 (SD = 0.54), respectively, implying that users did not perceive the system as overly complex or inconsistent.
These findings are further corroborated by user interviews. 
Eight users reported that the system was ``easy to navigate” and ``provided useful suggestions” during the annotation process.
User 7 stated, ``The interface was straightforward, and I felt confident in using the system even with limited art knowledge.”
However, a few users (users 4 and 5), particularly those with less experience in annotation, mentioned initial confusion with the clustering feature, although they quickly adapted.

\begin{figure}[h]
    \centering
    \includegraphics[width=\columnwidth]{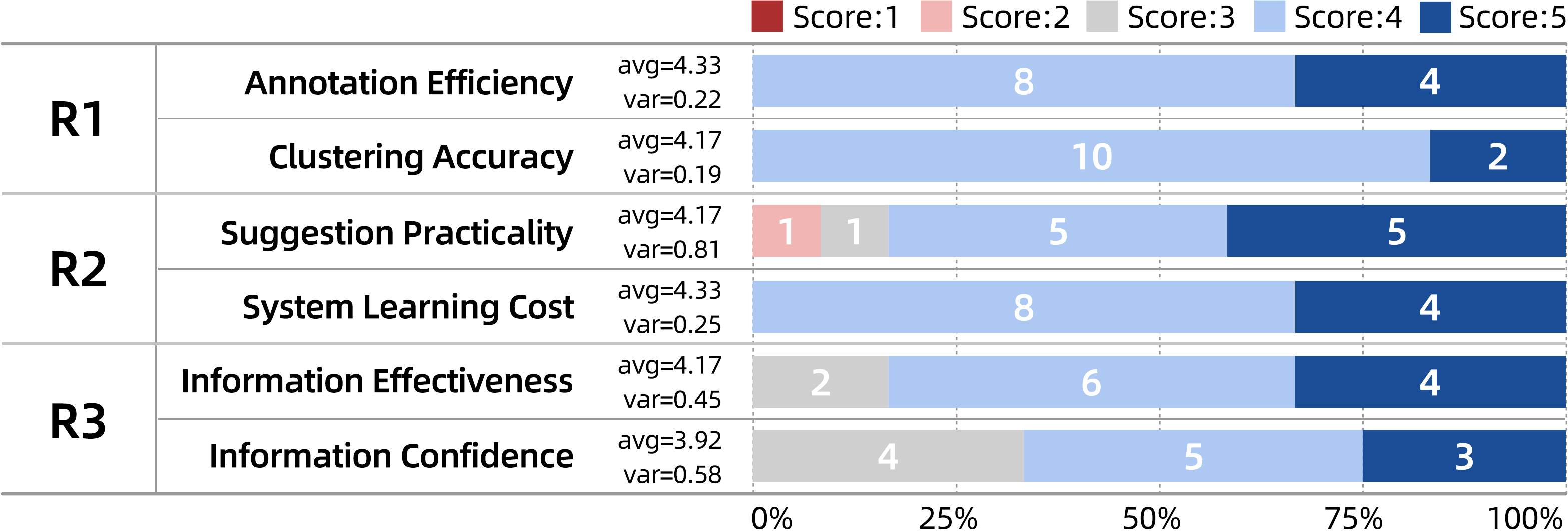}
    \caption{Results of the system requirement questionnaire.}
    \label{fig:eval}
\end{figure}

\textit{Task-Specific Evaluation.}
To evaluate whether the system meets the predefined design requirements, we conducted a task-specific evaluation using a 5-point Likert scale.
Six questions were designed to correspond to the design goals (R1–R3), and the detailed scores are shown in \cref{fig:eval}. 
The results indicate that the clustering feature significantly improves annotation efficiency, with an average score of 4.33 for Q1 and 4.17 for Q2. 
The accuracy of the clustering results was also rated high (M = 4.17), indicating that the clustering generally meets user expectations.
Regarding proactive assistance (R2), participants reported that the system offers useful suggestions at appropriate times (M = 4.17, Q3) and reduces the burden of manual exploration (M = 4.33, Q4). 
This indicates that the system effectively assists users during the annotation process and reduces cognitive load.
For interpretability and trust (R3), the system’s visual, textual, and knowledge-based support played a key role in helping users evaluate AI suggestions.
Q5 received a mean score of 4.17, indicating that the provided information assisted users in determining the correctness of AI suggestions, while Q6 (M = 3.92) showed a moderate increase in users' trust in AI recommendations. 
The lower ratings from 4 users (M = 3) suggest that those who already trusted the AI outputs felt less need for additional evidence, indicating that the system’s evidence mainly reinforced their existing trust rather than substantially increasing it.
In summary, these results suggest that the system meets its design goals for clustering support, proactive assistance, and interpretability.

\textit{Efficiency.}
We compared the annotation time and label agreement between the
baseline and our system across the three conditions.
For annotation time, our system substantially reduced the time cost
(M = 15.75 minutes) compared to the baseline (M = 30.92 minutes),
nearly a 50\% improvement ($p = .00049$). 
In terms of label agreement, our system achieved higher average
agreement (27/30, 90\%) than the baseline (22/30, 73\%)
($p = .00049$).
The reduction in time can be attributed to several factors observed
from user behavior.
First, users no longer needed to switch between the system and
external resources or repeatedly query AI, which reduced
task-switching overhead.
Second, proactive label recommendations helped users quickly narrow
down relevant labels, reducing the effort required to explore and
eliminate irrelevant options.
Third, cluster-based batch annotation enabled users to process similar
images collectively with less context switch.
Several participants reported improved efficiency with the system's
assistance; as User 1 noted, ``the recommendations and clustering
significantly reduce the time spent deciding what to annotate.”
Overall, these results demonstrate that the system not only improves
label agreement but also substantially reduces annotation time.

\textit{System Evolution.}
To evaluate whether the system supports reusable experience (R4), we
conducted an ablation study by enabling and disabling the
Interaction-Driven Evolution Module (C2 vs.\ C3).
In terms of efficiency, average annotation time decreased from 17.25
minutes (C2, without) to 15.75 minutes (C3, with), though this
difference was not statistically significant ($p = .088$).
The clearer benefit appeared in user experience: C3 was rated
significantly higher than C2 ($p = .00049$), and 7 of 12 participants
explicitly reported experience improvements attributable to the
evolution module. The non-significance in time may partly reflect the
small sample size, which we plan to address with larger-scale
validation.
This suggests that accumulated knowledge and skills can positively
support subsequent annotation tasks.
Quantitatively, the Behavior Mining Agent generated 112 raw
candidate skills during the study, of which 31 were retained after
structure-aware merging and redundancy removal---a 72\% reduction---
indicating that the module actively consolidates overlapping skills
rather than accumulating them unboundedly.
Moreover, most participants reported no additional operational burden when the module was enabled, while the system’s knowledge and recommendation rationales became more reasonable and detail ed, reducing the need to consult the chat interface. As User 10 noted, ``I didn’t feel any difference in how I operated the system, but it did seem to become smarter.”
Several users observed progressive improvement in recommendation quality: users who corrected or questioned the system found that earlier errors did not reappear, suggesting that the system learned from their feedback, while other users, even without providing corrections, benefited from previously accumulated skills and reported higher confidence in the system’s suggestions.
Overall, the accumulated knowledge improved system performance and user efficiency
without requiring additional user effort, demonstrating iterative skill
refinement across interaction rounds within a session.

\textit{Insights from interviews.}
The interviews revealed several key insights regarding Human-Agent interaction:
\textit{Efficiency and focus}: Proactive suggestions, guided clusters, and label recommendations reduced blind searching and helped users stay focused on annotation.
\textit{Understanding and learning}: Background knowledge and evidence-based cues supported interpretation, especially for users with limited art expertise; when recommendations conflicted with expectations, users often verified the information, sometimes leading to deeper reflection.
\textit{Trust and concerns}: While participants valued interaction-driven system improvement and knowledge accumulation, some expressed concerns about over-reliance on recommendations and the need to consider data privacy.

\subsection{Case Study}
We evaluate our system through two cases sourced from real data annotation tasks.
The first, \textit{Annotation of Female Identity in Traditional Chinese Paintings}, illustrates how Agent augments human annotation performance and structures expert annotation trajectories into reusable system knowledge. 
The second, \textit{Annotation of Implicit Meanings in Mexican Posters}, demonstrates how previously codified experiences are invoked, transferred, and adapted to new contextual demands.

\subsubsection{Case 1: Annotation of Female Identity in Traditional Chinese Paintings}
In this case, we illustrate the system with a female-identity
annotation task in Traditional Chinese Painting, labeling roles such
as maids, court ladies, or empresses, to collect data for future
research on the visual representation and social identity of women.
User 6, an experienced annotator with limited art knowledge, used
\systemname to complete it, assigning each figure a label from a
predefined taxonomy.
At the beginning, the Skill Recommendation Agent suggested three relevant skills, and the user selected female figure identity classification and multi-person relationship recognition  (\cref{fig:teaser}(C1)). 
The Cluster and Pattern Mining Agent then grouped the images by
visual and semantic similarity and extracted cluster-level patterns,
such as marginal standing positions, forward-leaning postures, and
object-holding gestures (\cref{fig:teaser}(F1)).
Based on these, the Label Recommendation Agent aligned them with the label schema and recommended the cluster.

The system highlighted the cluster and presented a pop-up label recommendation and explanation (\cref{fig:teaser}(C2)).
After reviewing the recommended labels and the supporting evidence---
cluster recommendations~(\cref{fig:teaser}(D1)), background
knowledge~(\cref{fig:teaser}(D2)), and visual
patterns~(\cref{fig:teaser}(D3))---User 6 noticed a potential error
in the recommendation for \textit{The Third Panel of Four-Season
Beauties}.
Unlike other images in the cluster, the two figures in this image showed no clear service-related actions, and their clothing appeared similar, suggesting equal status rather than a master–servant relationship.
To verify this, User 6 queried the system: \textit{whether there are cases in traditional Chinese paintings where front–back arrangements do not imply subordination.} 
The system retrieved additional examples from related clusters and provided comparative evidence (\cref{fig:teaser}(D)).
Based on this, User 6 confirmed that the two figures were of equal status and labeled them as ``female companions.”

Instead of treating this as a simple correction, the system recorded the full interaction and extracted a key principle: spatial position is a weak cue, while functional actions are stronger indicators of social roles.
The principle was stored in the knowledge base and translated into
executable skills. A domain-specific skill, \textit{attendant
recognition in multi-figure paintings}, guides the system to inspect
hand gestures, object interactions, and posture with zoom-in tools,
compare clothing, and infer roles. A more general skill,
\textit{spatial relations do not imply semantic subordination},
enforces checking behavioral evidence before assigning roles.
Principle and skill examples are provided in the supplementary material.

In a later stage, user 8 encountered a similar cluster containing \textit{Ladies Picking Flowers}.
Guided by the learned skill, user 8 followed the recommended analysis process by examining hand gestures, object interactions, and clothing details instead of relying solely on spatial arrangement.
As a result, although the figures were also arranged in a front–back composition, the absence of service-oriented actions and the similarity in appearance led to a correct interpretation of equal status, rather than misclassifying the rear figure as an attendant.
This case demonstrates bidirectional Human-AI assistance: the system supports users during annotation, while user feedback continuously improves the system by transforming experience into reusable knowledge and skills, which can be leveraged in subsequent tasks to improve overall efficiency.
\begin{figure}
    \centering
    \includegraphics[width=1\linewidth]{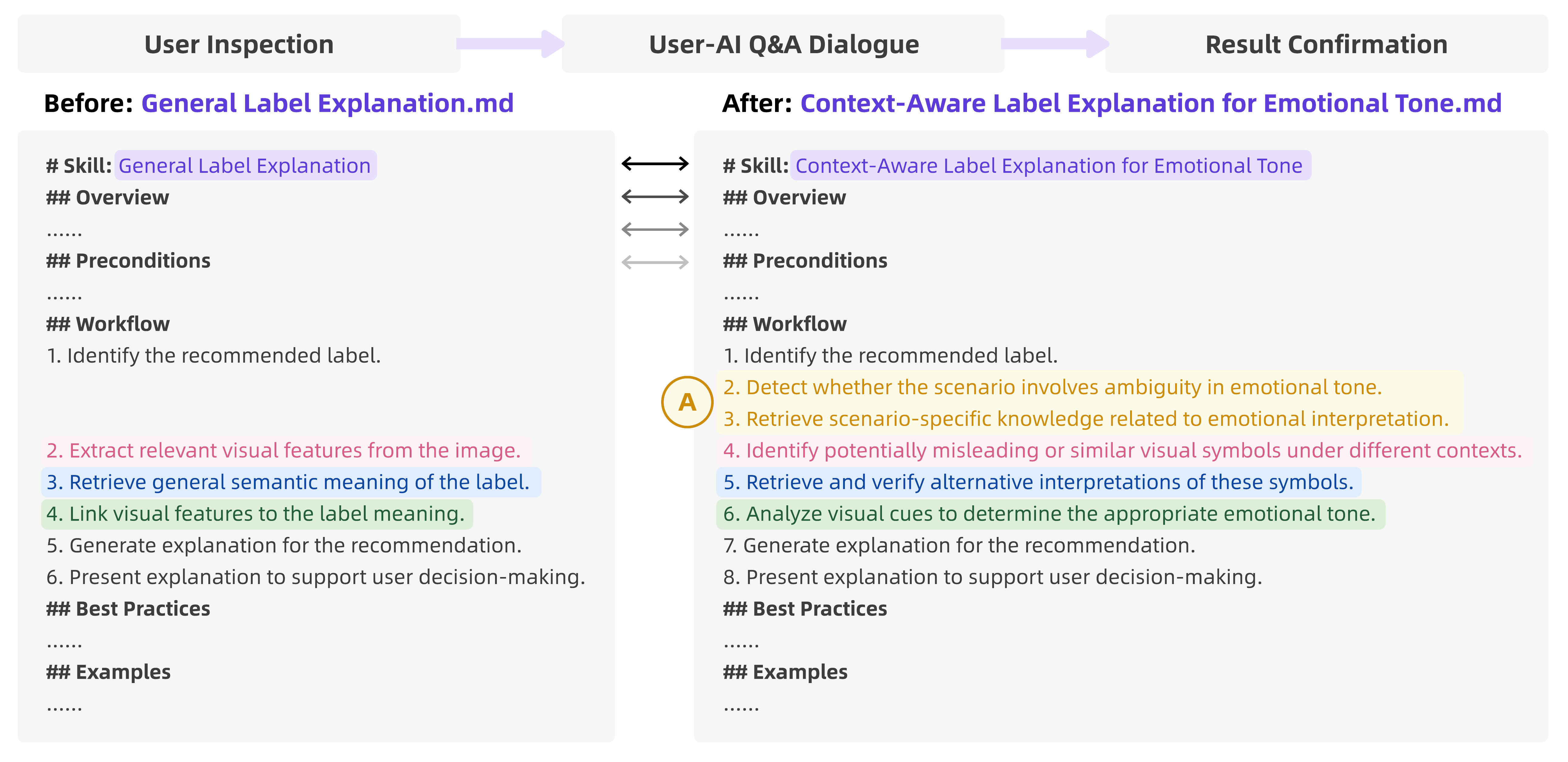}
    \caption{Comparison of skills before and after annotation. The left shows the pre-annotation skill, while the right shows the refined post-annotation skill. (A) highlights the key differences between the two.}
    \label{Fig:case2}
\end{figure}
\subsubsection{Case 2: Annotation of Implicit Meanings in Mexican Posters for Knowledge Transfer and Adaptation}

This case is drawn from an emotional-tone annotation task on Mexican
posters, supporting research on cross-cultural understanding and
translation. While annotating a poster, the user received a
recommendation labeling a skull element as ``celebration,'' with an
explanation grounded in local culture: in traditions such as
\textit{Day of the Dead}, skull imagery honors the dead and bright
colors reinforce a festive atmosphere. The user found this confusing,
since in Western contexts skulls are often tied to Halloween, death,
or fear. After the user questioned this, the system provided a
contrastive explanation of skull symbolism across Mexican and Western
contexts, and the user accepted the label.
This interaction drives a refinement of the skill: rather than giving
a single-culture justification, the evolved skill proactively
anticipates cross-cultural misunderstandings by retrieving and
contrasting alternative interpretations (e.g., Western associations of
skulls with horror) and explaining why the recommended label remains
valid in context~(\cref{Fig:case2}). The knowledge base is updated to
encode this contrastive reasoning pattern, making future explanations
more robust and context-aware without extra clarification turns.
This shows that a learned skill can generalize across scenarios and
adapt to new tasks.

\section{Discussion}
We discuss our work from three key aspects: the effectiveness and generalizability of the system, the reflection of Human-Agent collaboration and co-evolution, and the limitations and future directions.

\subsection{System Effectiveness and Generalizability}
Our results show that \systemname improves artworks annotation effectiveness by clustering similar cases for grouped comparison, proactively recommending relevant labels and clusters, providing multimodal evidence to support verification, and turning users’ questions and corrections into reusable skills and knowledge base.
By providing contextual guidance and interpretable evidence, the
system also reduces the reliance on prior domain expertise.
Through our bidirectional augmentation, the system not only improves the efficiency of finishing tasks but also iteratively refines its capabilities across interaction rounds by incorporating skills and knowledge learned from human users. 

More broadly, we expect our approach to extend beyond artwork annotation to
other knowledge-intensive annotation scenarios, such as medical image annotation,
pathology image analysis, archaeological image documentation, and historical
document image organization, where domain expertise is essential for both
interpretation and verification.
We expect the behavior mining pipeline, and the
Interaction-Driven Evolution Module to be directly transferable, as they are not
tied to artwork-specific semantics; in contrast, the labeling taxonomy, the domain
knowledge base, and task-specific prompts would require substantial domain
adaptation.
For example, in pathology image annotation, the labeling taxonomy would need to be
redesigned around lesion types, tissue structures, and disease stages, with medical
knowledge and diagnostic criteria incorporated into the knowledge base.
We note, however, that our current evidence comes solely from artwork annotation,
and validating this transferability in other domains remains future work.

\subsection{Human-AI Collaboration and Co-Evolution}
The bidirectional design of the \textit{\frameworkname} framework extends beyond annotation, offering a robust template for any Human-AI collaboration that seeks to utilize AI to augment human productivity and transform transient interactions into enduring, reusable machine intelligence.
Rather than treating AI merely as a passive command executors, our framework proposes a bidirectional augmentation paradigm: while the AI proactively empowers human efficiency in completing tasks, the system simultaneously internalizes human intelligence by distilling decision rationales and correction processes into reusable knowledge and skills. 
This ensures that human expertise is not just consumed, but serves as the catalyst for the continuous, recursive evolution of the model's capabilities. 
For example, in medical diagnosis support, AI can provide candidate diagnoses and retrieve similar cases, while doctors make judgments, corrections, and provide additional reasoning.
These decision processes and corrections can be recorded and transformed into new diagnostic knowledge and reasoning strategies, which can then support future cases and other practitioners. 

This framework also raises a broader discussion on Human-AI coevolution~\cite{coevolution,evoAgents,coevolution2}.
While current research often focuses on autonomous AI evolution toward general intelligence, our work suggests an alternative trajectory: a reciprocal evolutionary process where human expertise and machine intelligence grow in tandem. 
Instead of viewing AI as a self-contained evolving entity, we argue for a paradigm where the interaction itself serves as the engine for collective intelligence. 
For example, in educational settings, AI can act as an instructor, accelerating learners' skill acquisition by providing explanations, feedback, and exercises, while learners’ problem-solving processes and feedback can be captured to improve the AI’s teaching strategies and knowledge organization.
In this way, human learning and AI capability improvement can form a mutually reinforcing loop.

\subsection{Limitations and Future Work}

While the proposed system offers a robust approach for collaborative Human-AI annotation tasks, several limitations exist that should be addressed in future iterations to enhance its adaptability, efficiency, and performance.

\textit{Lack of Personalization and Flexibility in User Interaction.}
The current system adopts a proactive recommendation approach, which structures the order and grouping of images. While this design benefits a group of users, others may prefer to browse images according to their own pace and workflow. 
This would make the system more adaptive to individual user needs and usage habits, enhancing the overall user experience by providing a more personalized and responsive interaction model.

\textit{Skill Reliability and Annotator Bias.}
The system relies on preset algorithms (\cref{alg:skill_merge}) and
the agent's autonomous decisions to update and merge skills, but does
not yet validate annotation quality.
Since interaction sequences carry no correctness markers, noisy
actions such as mislabeling may propagate into learned skills, while
conflicting interpretations across users are intentionally retained to
preserve interpretive diversity. Future versions should add quality
assurance, e.g., expert review of generated skills.
Skills in ArtAnno are distilled from human annotations, so the
system inherits its annotators' biases. In the Mexican-poster case,
the system defaulted to a Western ``Halloween'' reading, corrected
only because the user noticed the anomaly; had it gone unnoticed, the
biased skill would have been retained and reused, propagating the
error.
Future work should therefore add bias detection,
per-skill confidence estimation, and provenance tracking that records
where each skill originates and gates its reuse through expert
review.

\textit{Limited Flexibility in Annotation Framework for Multiple Labels.}
The current system uses a relatively fixed annotation framework, which cannot handle situations where a single object requires multiple labels.
For example, in Traditional Chinese Painting, symbols and elements may have multiple interpretations. 
Future versions of the system should allow for more flexible and adaptable tagging of objects, such as multi-label annotation, thus better accommodating diverse art forms and contexts.

\textit{Lack of Large-Scale, Longitudinal Testing.}
The current system has been evaluated only in controlled in-lab environments. 
Future work should conduct longitudinal evaluations with diverse users and larger datasets to assess scalability, usability, and long-term reliability, particularly how the skill and knowledge base evolves and adapts over extended use.

\section{Conclusion}
In this paper, we address the persistent challenge of efficient and knowledge-intensive artwork annotation by proposing \textit{BiHAA}, a bidirectional Human-AI assistance framework that unifies AI-supported task completion with human-interaction-driven system evolution.
Guided by formative findings, we designed and implemented ArtAnno, a multi-agent system that provides proactive assistance while continuously capturing and reusing human expertise through a Skill Library and Knowledge Base.
Our results from user study and case study show that ArtAnno not only improves annotation efficiency 
but also enables the accumulation and reuse of both domain knowledge
and procedural strategies across interaction rounds within a
session.
More broadly, this work suggests that artwork annotation systems should move beyond one-way assistance toward a collaborative paradigm in which humans and AI continuously enhance one another. 

\balance

\begin{acks}
This work was supported by the National Natural Science Foundation of China under Grant 62502423 and Grant 62421003.
This work involved human subjects in its research. Approval of all ethical and experimental procedures and protocols was granted by the Ethics Committee of the College of Biomedical Engineering and Instrument Science, Zhejiang University under Application No. [2025]36.
\end{acks}

\bibliographystyle{ACM-Reference-Format}
\bibliography{reference, intro, relatedWork}

@article{shi2026survey,
  title={A Survey of Human-AI Collaboration for Scientific Discovery},
  author={Shi, Chuhan and Ren, Xiaoquan and Wang, Yifang and Li, Junze and Sun, Yushi and Luo, Yawen and Sheng, Rui},
  year={2026},
  url={https://doi.org/10.20944/preprints202601.0405.v2}
}

@article{wang2026figures,
  title={Figures as Interfaces: Toward LLM-Native Artifacts for Scientific Discovery},
  author={Wang, Yifang and Sheng, Rui and Shao, Erzhuo and Qian, Yifan and Li, Haotian and Cao, Nan and Wang, Dashun},
  journal={arXiv preprint arXiv:2604.08491},
  year={2026},
  url={https://doi.org/10.48550/arXiv.2604.08491}
}

@article{shao2026sciscigpt,
  title={SciSciGPT: Advancing Human--AI Collaboration in the Science of Science},
  author={Shao, Erzhuo and Wang, Yifang and Qian, Yifan and Pan, Zhenyu and Liu, Han and Wang, Dashun},
  journal={Nature Computational Science},
  volume={6},
  number={3},
  pages={301--315},
  year={2026},
  url={https://doi.org/10.1038/s43588-025-00906-6}
}

@inproceedings{DeArt,
author = {Reshetnikov, Artem and Marinescu, Maria-Cristina and Lopez, Joaquim More},
title = {DEArt: Dataset of European Art},
year = {2022},
isbn = {978-3-031-25055-2},
url = {https://doi.org/10.1007/978-3-031-25056-9_15},
doi = {10.1007/978-3-031-25056-9_15},
booktitle = {Computer Vision – ECCV Workshops: Tel Aviv, Israel, Proceedings, Part I},
pages = {218--233},
numpages = {16},
location = {Tel Aviv, Israel}
}

@misc{artbench,
doi = {10.21227/typ6-a328},
url = {https://dx.doi.org/10.21227/typ6-a328},
author = {Ravidu Suien Rammuni Silva},
title = {AI-ArtBench},
year = {2025} }

@misc{artEmis,
      title={ArtEmis: Affective Language for Visual Art}, 
      author={Panos Achlioptas and Maks Ovsjanikov and Kilichbek Haydarov and Mohamed Elhoseiny and Leonidas Guibas},
      year={2021},
      eprint={2101.07396},
      archivePrefix={arXiv},
      primaryClass={cs.CV},
      url={https://arxiv.org/abs/2101.07396}, 
}

@inproceedings{lin2014coco,
  author = {Tsung-Yi Lin and Michael Maire and Serge Belongie and James Hays and Pietro Perona and Deva Ramanan and Piotr Dollár and C. Lawrence Zitnick},
  title = {Microsoft COCO: Common Objects in Context},
  booktitle = {European Conference on Computer Vision (ECCV)},
  pages = {740--755},
  year = {2014},
  url = {https://doi.org/10.1007/978-3-319-10602-1_48}
}

@ARTICLE{markup,
AUTHOR={Dobbie, Samuel  and Strafford, Huw  and Pickrell, W. Owen  and Fonferko-Shadrach, Beata  and Jones, Carys  and Akbari, Ashley  and Thompson, Simon  and Lacey, Arron },
TITLE={Markup: A Web-Based Annotation Tool Powered by Active Learning},
JOURNAL={Frontiers in Digital Health},
VOLUME={Volume 3 - 2021},
YEAR={2021},
URL={https://www.frontiersin.org/journals/digital-health/articles/10.3389/fdgth.2021.598916},
DOI={10.3389/fdgth.2021.598916},
ISSN={2673-253X},
}

@article{semianno,
author = {Benato, B\'{a}rbara C. and Gomes, Jancarlo F. and Telea, Alexandru C. and Falc\~{a}o, Alexandre X.},
title = {Semi-Automatic Data Annotation Guided by Feature Space Projection},
year = {2021},
volume = {109},
number = {C},
issn = {0031-3203},
url = {https://doi.org/10.1016/j.patcog.2020.107612},
doi = {10.1016/j.patcog.2020.107612},
journal = {Pattern Recogn.},
numpages = {11},
}

@String{Computing = "Computing" }

@String{Computer = "{IEEE} Computer" }

@String{Academic = "Academic Press" }

@String{Macmillan = "Macmillan" }

@inproceedings{cheng2024yolo,
  title={YOLO-World: Real-Time Open-Vocabulary Object Detection},
  author={Cheng, Tianheng and Song, Lin and Ge, Yixiao and Liu, Wenyu and Wang, Xinggang and Shan, Ying},
  booktitle={Proceedings of the IEEE/CVF conference on computer vision and pattern recognition},
  pages={16901--16911},
  year={2024},
  url = {https://arxiv.org/abs/2401.17270}
}

@inproceedings{postermate,
author = {Shin, Donghoon and Lee, Daniel and Hsieh, Gary and Chan, Gromit Yeuk-Yin},
title = {PosterMate: Audience-driven Collaborative Persona Agents for Poster Design},
year = {2025},
isbn = {9798400720376},
url = {https://doi.org/10.1145/3746059.3747769},
doi = {10.1145/3746059.3747769},
booktitle = {Proceedings of the 38th Annual ACM Symposium on User Interface Software and Technology},
articleno = {201},
numpages = {20},
series = {UIST}
}

@ARTICLE{vista,
  author={Xuan, Xiwei and Wang, Xiaoqi and He, Wenbin and Ono, Jorge Piazentin and Gou, Liang and Ma, Kwan-Liu and Ren, Liu},
  journal={IEEE Transactions on Visualization and Computer Graphics}, 
  title={VISTA: A Visual Analytics Framework to Enhance Foundation Model-Generated Data Labels}, 
  year={2025},
  volume={31},
  number={10},
  pages={6991--7003},
  doi={10.1109/TVCG.2025.3535896}
}

@inproceedings{coevolution,
    title = "Human-{LLM} Coevolution: Evidence from Academic Writing",
    author = "Geng, Mingmeng  and
      Trotta, Roberto",
    editor = "Che, Wanxiang  and
      Nabende, Joyce  and
      Shutova, Ekaterina  and
      Pilehvar, Mohammad Taher",
    booktitle = "Findings of the Association for Computational Linguistics (ACL)",
    year = "2025",
    address = "Vienna, Austria",
    url = "https://aclanthology.org/2025.findings-acl.657/",
    doi = "10.18653/v1/2025.findings-acl.657",
    pages = "12689--12696",
    ISBN = "979-8-89176-256-5",
}

@inproceedings{coevolution2,
  author       = {Nyasha Kadenhe and Mohamed Al Musleh and Allan Lompot},
  title        = {Human-AI Co-Design and Co-Creation: A Review of Emerging Approaches, Challenges, and Future Directions},
  booktitle    = {Proceedings of the 2025 AAAI Summer Symposium Series: Human-AI Collaboration: Exploring Diversity of Human Cognitive Abilities and Varied AI Models for Hybrid Intelligent Systems},
  volume       = {6},
  number       = {1},
  year         = {2025},
  doi          = {10.1609/aaaiss.v6i1.36061},
  url          = {https://doi.org/10.1609/aaaiss.v6i1.36061},
}

@inproceedings{multiagentmultimodal,
author = {Perera, Madhawa and Hossain, Md Zakir and Krumpholz, Alexander and Gedeon, Tom},
title = {Developing Multimodal Human-AI Interaction Systems Using Multi-Agent Frameworks},
year = {2025},
isbn = {9798400720765},
url = {https://doi.org/10.1145/3747327.3762827},
doi = {10.1145/3747327.3762827},
booktitle = {Companion Proceedings of the 27th International Conference on Multimodal Interaction},
pages = {67--69},
numpages = {3},
series = {ICMI Companion}
}

@inproceedings{medbiaser,
author = {Shi, Shaohan and Shao, Yuheng and Jiang, Haoran and Yao, Yunjie and Zhang, Zhijun and Ding, Xu and Li, Quan},
title = {MEDebiaser: A Human-AI Feedback System for Mitigating Bias in Multi-label Medical Image Classification},
year = {2025},
isbn = {9798400720376},
url = {https://doi.org/10.1145/3746059.3747725},
doi = {10.1145/3746059.3747725},
booktitle = {Proceedings of the 38th Annual ACM Symposium on User Interface Software and Technology},
articleno = {180},
numpages = {27},
series = {UIST}
}

@inproceedings{KALE,
author = {Jiang, Yanbei and Ehinger, Krista A. and Lau, Jey Han},
title = {KALE: An Artwork Image Captioning System Augmented with Heterogeneous Graph},
year = {2024},
isbn = {978-1-956792-04-1},
url = {https://doi.org/10.24963/ijcai.2024/848},
doi = {10.24963/ijcai.2024/848},
booktitle = {Proceedings of the Thirty-Third International Joint Conference on Artificial Intelligence},
articleno = {848},
numpages = {9},
location = {Jeju, Korea},
series = {IJCAI}
}

@article{cataanno,
author = {Shao, Hanning and Yuan, Xiaoru},
title = {CataAnno: An Ancient Catalog Annotator for Annotation Cleaning by Recommendation},
year = {2025},
issue_date = {Jan. 2025},
volume = {31},
number = {1},
issn = {1077-2626},
url = {https://doi.org/10.1109/TVCG.2024.3456379},
doi = {10.1109/TVCG.2024.3456379},
pages = {404--414},
numpages = {11}
}

@inproceedings{humanLLMcoannotation,
author = {Wang, Xinru and Kim, Hannah and Rahman, Sajjadur and Mitra, Kushan and Miao, Zhengjie},
title = {Human-LLM Collaborative Annotation Through Effective Verification of LLM Labels},
year = {2024},
isbn = {9798400703300},
url = {https://doi.org/10.1145/3613904.3641960},
doi = {10.1145/3613904.3641960},
booktitle = {Proceedings of the CHI Conference on Human Factors in Computing Systems},
articleno = {303},
numpages = {21},
series = {CHI}
}

@article{MedSAM,
  author  = {Jun Ma and Yuting He and Feifei Li and Lin Han and Chenyu You and Bo Wang},
  title   = {Segment Anything in Medical Images},
  journal = {Nature Communications},
  year    = {2024},
  volume  = {15},
  pages   = {654},
  doi     = {10.1038/s41467-024-44824-z},
  url     = {https://doi.org/10.1038/s41467-024-44824-z}
}

@article{SemanticAnno,
author = {Uren, Victoria and Cimiano, Philipp and Iria, Jos\'{e} and Handschuh, Siegfried and Vargas-Vera, Maria and Motta, Enrico and Ciravegna, Fabio},
title = {Semantic Annotation for Knowledge Management: Requirements and a Survey of the State of the Art},
year = {2006},
address = {NLD},
volume = {4},
number = {1},
issn = {1570-8268},
url = {https://doi.org/10.1016/j.websem.2005.10.002},
doi = {10.1016/j.websem.2005.10.002},
journal = {Web Semant.},
month = jan,
pages = {14--28},
numpages = {15},
}

@ARTICLE{CalliVA,
  author={Li, Jincheng and Wu, Jinpeng and Tan, Shaocong and Du, Lin and Zhang, Yu and Yang, Chaofan and Zhang, Jiadi and Xu, Rebecca Ruige and Shi, Rui and Bai, Lu and Yuan, Xiaoru},
  journal={IEEE Transactions on Visualization and Computer Graphics}, 
  title={Calli-VA: A Visual Analytics System for Analyzing and Comparing Chinese Calligraphic Styles}, 
  year={2026},
  volume={32},
  number={1},
  pages={955--965},
  doi={10.1109/TVCG.2025.3634633}}

@inproceedings{creativityAgent,
author = {Liu, Yiren and Chen, Si and Cheng, Haocong and Yu, Mengxia and Ran, Xiao and Mo, Andrew and Tang, Yiliu and Huang, Yun},
title = {How AI Processing Delays Foster Creativity: Exploring Research Question Co-Creation with an LLM-based Agent},
year = {2024},
isbn = {9798400703300},
url = {https://doi.org/10.1145/3613904.3642698},
doi = {10.1145/3613904.3642698},
booktitle = {Proceedings of the 2024 CHI Conference on Human Factors in Computing Systems},
articleno = {17},
numpages = {25},
series = {CHI}
}

@inproceedings{IAI,
  author    = {Leixian Shen and Yifang Wang and Huamin Qu and Xing Xie and Haotian Li},
  title     = {Interaction-Augmented Instruction: Modeling the Synergy of Prompts and Interactions in Human-GenAI Collaboration},
  booktitle = {Proceedings of the {CHI} Conference on Human Factors in Computing Systems},
  year      = {2026},
  doi       = {10.1145/3772318.3790505},
  url       = {https://www.microsoft.com/en-us/research/publication/interaction-augmented-instruction-modeling-the-synergy-of-prompts-and-interactions-in-human-genai-collaboration/}
}

@article{PAMSNet,
title = {PAMSNet: A Point Annotation-Driven Multi-Source Network for Remote Sensing Semantic Segmentation},
journal = {ISPRS Journal of Photogrammetry and Remote Sensing},
volume = {229},
pages = {1--16},
year = {2025},
issn = {0924-2716},
doi = {https://doi.org/10.1016/j.isprsjprs.2025.07.035},
url = {https://www.sciencedirect.com/science/article/pii/S0924271625003041},
author = {Yuanhao Zhao and Mingming Jia and Genyun Sun and Aizhu Zhang},
}

@ARTICLE{KMTLabeler,
  author={Wang, He and Ouyang, Yang and Wu, Yuchen and Jiang, Chang and Jin, Lixia and Cao, Yuanwu and Li, Quan},
  journal={IEEE Transactions on Visualization and Computer Graphics}, 
  title={KMTLabeler: An Interactive Knowledge-Assisted Labeling Tool for Medical Text Classification}, 
  year={2025},
  volume={31},
  number={9},
  pages={4493--4510},
  doi={10.1109/TVCG.2024.3406387}}

@ARTICLE{coevolution3,
AUTHOR={Högberg, Anders },
TITLE={Becoming Human in the Age of AI: Cognitive Co-Evolutionary Processes},
JOURNAL={Frontiers in Psychology},
VOLUME={Volume 16 - 2025},
YEAR={2026},
URL={https://www.frontiersin.org/journals/psychology/articles/10.3389/fpsyg.2025.1734048},
DOI={10.3389/fpsyg.2025.1734048},
ISSN={1664-1078},
}

@INPROCEEDINGS{visNarra,
  author={Springstein, Matthias and Schneider, Stefanie and Rahnama, Javad and Stalter, Julian and Kristen, Maximilian and Müller-Budack, Eric and Ewerth, Ralph},
  booktitle={IEEE/CVF Winter Conference on Applications of Computer Vision (WACV)}, 
  title={Visual Narratives: Large-scale Hierarchical Classification of Art-historical Images}, 
  year={2024},
  pages={7195--7205},
  doi={10.1109/WACV57701.2024.00705}}

@article{Dubourg2024CulturalAnnotation,
  author  = {Emilie Dubourg and Vincent Thouzeau and Nicolas Baumard},
  title   = {A Step-by-Step Method for Cultural Annotation by LLMs},
  journal = {Frontiers in Artificial Intelligence},
  year    = {2024},
  volume  = {7},
  pages   = {1365508},
  doi     = {10.3389/frai.2024.1365508},
  pmid    = {38756758},
  pmcid   = {PMC11097685}
}

@misc{softwareAgent,
      title={Understanding Software Engineering Agents: A Study of Thought-Action-Result Trajectories}, 
      author={Islem Bouzenia and Michael Pradel},
      year={2025},
      eprint={2506.18824},
      archivePrefix={arXiv},
      primaryClass={cs.SE},
      url={https://arxiv.org/abs/2506.18824}, 
}

@article{decisionAgent,
title = {Can an AI Agent Lead Human Teams?},
journal = {Computers in Human Behavior: Artificial Humans},
volume = {7},
pages = {100278},
year = {2026},
issn = {2949-8821},
doi = {https://doi.org/10.1016/j.chbah.2026.100278},
url = {https://www.sciencedirect.com/science/article/pii/S2949882126000290},
author = {James Simpson and Gaurav Patil and Hamish Stening and Ayman {Bin Kamruddin} and Daniel Somerville and Sigrid Seage and Patrick Nalepka and Mark Dras and Simon G. Hosking and Rachel W. Kallen and Michael J. Richardson and Deborah Richards},
}

@misc{xskill,
      title={XSkill: Continual Learning from Experience and Skills in Multimodal Agents}, 
      author={Guanyu Jiang and Zhaochen Su and Xiaoye Qu and Yi R. Fung},
      year={2026},
      eprint={2603.12056},
      archivePrefix={arXiv},
      primaryClass={cs.AI},
      url={https://arxiv.org/abs/2603.12056}, 
}

@misc{autoskill,
      title={AutoSkill: Experience-Driven Lifelong Learning via Skill Self-Evolution}, 
      author={Yutao Yang and Junsong Li and Qianjun Pan and Bihao Zhan and Yuxuan Cai and Lin Du and Jie Zhou and Kai Chen and Qin Chen and Xin Li and Bo Zhang and Liang He},
      year={2026},
      eprint={2603.01145},
      archivePrefix={arXiv},
      primaryClass={cs.AI},
      url={https://arxiv.org/abs/2603.01145}, 
}

@misc{ModelingCollaborator,
      title={Modeling Collaborator: Enabling Subjective Vision Classification With Minimal Human Effort via LLM Tool-Use}, 
      author={Imad Eddine Toubal and Aditya Avinash and Neil Gordon Alldrin and Jan Dlabal and Wenlei Zhou and Enming Luo and Otilia Stretcu and Hao Xiong and Chun-Ta Lu and Howard Zhou and Ranjay Krishna and Ariel Fuxman and Tom Duerig},
      year={2024},
      eprint={2403.02626},
      archivePrefix={arXiv},
      primaryClass={cs.CV},
      url={https://arxiv.org/abs/2403.02626}, 
}

@article{evoAgents,
  author  = {Zhishang Xiang and Chengyi Yang and Zerui Chen and others},
  title   = {A Systematic Survey of Self-Evolving Agents: From Model-Centric to Environment-Driven Co-Evolution},
  journal = {TechRxiv},
  year    = {2026},
  url     = {https://www.techrxiv.org/doi/pdf/10.36227/techrxiv.177203250.05832634/v2}
}

@misc{surveyselfevolvingagentswhat,
      title={A Survey of Self-Evolving Agents: What, When, How, and Where to Evolve on the Path to Artificial Super Intelligence}, 
      author={Huan-ang Gao and Jiayi Geng and Wenyue Hua and Mengkang Hu and Xinzhe Juan and Hongzhang Liu and Shilong Liu and Jiahao Qiu and Xuan Qi and Yiran Wu and Hongru Wang and Han Xiao and Yuhang Zhou and Shaokun Zhang and Jiayi Zhang and Jinyu Xiang and Yixiong Fang and Qiwen Zhao and Dongrui Liu and Qihan Ren and Cheng Qian and Zhenhailong Wang and Minda Hu and Huazheng Wang and Qingyun Wu and Heng Ji and Mengdi Wang},
      year={2026},
      eprint={2507.21046},
      archivePrefix={arXiv},
      primaryClass={cs.AI},
      url={https://arxiv.org/abs/2507.21046}, 
}

@misc{LabelStudio2026,
  title     = {{Label Studio}: Data Labeling Software},
  author    = {Maxim Tkachenko and Mikhail Malyuk and Andrey Holmanyuk and Nikolai Liubimov},
  year      = {2026},
  howpublished = {\url{https://github.com/heartexlabs/label-studio}},
}

@article{DeepEdit,
  title={DeepEdit: Knowledge Editing as Decoding with Constraints},
  author={Yiwei Wang and Muhao Chen and Nanyun Peng and Kai-Wei Chang},
  journal={ArXiv},
  year={2024},
  volume={abs/2401.10471},
  url={https://api.semanticscholar.org/CorpusID:267060897}
}

@misc{VIA_tool,
  author = {Abhishek Dutta and Andrew Zisserman},
  title  = {VGG Image Annotator (VIA)},
  year   = {2017},
  howpublished = {\url{https://www.robots.ox.ac.uk/~vgg/software/via/}},
  note   = {Accessed: 2026}
}

@misc{DuoDrama,
      title={DuoDrama: Supporting Screenplay Refinement Through LLM-Assisted Human Reflection}, 
      author={Yuying Tang and Xinyi Chen and Haotian Li and Xing Xie and Xiaojuan Ma and Huamin Qu},
      year={2026},
      eprint={2602.05854},
      archivePrefix={arXiv},
      primaryClass={cs.HC},
      url={https://arxiv.org/abs/2602.05854}, 
}

@misc{artseek,
      title={ArtSeek: Deep Artwork Understanding via Multimodal In-Context Reasoning and Late Interaction Retrieval}, 
      author={Nicola Fanelli and Gennaro Vessio and Giovanna Castellano},
      year={2025},
      eprint={2507.21917},
      archivePrefix={arXiv},
      primaryClass={cs.CV},
      url={https://arxiv.org/abs/2507.21917}, 
}

@misc{artimgannoframework,
      title={Knowledge Graph for Intelligent Generation of Artistic Image Creation: Constructing a New Annotation Hierarchy}, 
      author={Jia Kaixin and Zhu Kewen and Deng Huanghuang and Qiu Yiwu and Ding Shiying and Ding Chenyang and Ning Zou and Li Zejian},
      year={2025},
      eprint={2511.03585},
      archivePrefix={arXiv},
      primaryClass={cs.HC},
      url={https://arxiv.org/abs/2511.03585}, 
}

@book{berger1972ways,
  author    = {Berger, John},
  title     = {Ways of Seeing},
  publisher = {Penguin Books},
  year      = {1972},
  isbn      = {0-14-013515-4},
}

@book{mitchell1994picture,
  author    = {Mitchell, W. J. Thomas},
  title     = {Picture Theory},
  year      = {1994},
  publisher = {University of Chicago Press},
  edition   = {2},
  isbn      = {0226532313},
}

@inproceedings{eCul,
author = {Schreiber, Guus and Amin, Alia and van Assem, Mark and de Boer, Victor and Hardman, Lynda and Hildebrand, Michiel and Hollink, Laura and Huang, Zhisheng and van Kersen, Janneke and de Niet, Marco and Omelayenko, Borys and van Ossenbruggen, Jacco and Siebes, Ronny and Taekema, Jos and Wielemaker, Jan and Wielinga, Bob},
title = {MultimediaN E-Culture Demonstrator},
year = {2006},
isbn = {3540490299},
url = {https://doi.org/10.1007/11926078_70},
doi = {10.1007/11926078_70},
booktitle = {Proceedings of the 5th International Conference on The Semantic Web},
pages = {951--958},
numpages = {8},
series = {ISWC}
}

@misc{CVAT2024,
  title        = {Computer Vision Annotation Tool (CVAT)},
  author       = {{CVAT.ai Corporation}},
  year         = {2024},
  howpublished = {\url{https://github.com/cvat-ai/cvat}},
  doi          = {10.5281/zenodo.10977499}
}

@misc{medsam2,
      title={MedSAM2: Segment Anything in 3D Medical Images and Videos}, 
      author={Jun Ma and Zongxin Yang and Sumin Kim and Bihui Chen and Mohammed Baharoon and Adibvafa Fallahpour and Reza Asakereh and Hongwei Lyu and Bo Wang},
      year={2025},
      eprint={2504.03600},
      archivePrefix={arXiv},
      primaryClass={eess.IV},
      url={https://arxiv.org/abs/2504.03600}, 
}

@misc{crowdagent,
      title={CrowdAgent: Multi-Agent Managed Multi-Source Annotation System}, 
      author={Maosheng Qin and Renyu Zhu and Mingxuan Xia and Chenkai Chen and Zhen Zhu and Minmin Lin and Junbo Zhao and Lu Xu and Changjie Fan and Runze Wu and Haobo Wang},
      year={2025},
      eprint={2509.14030},
      archivePrefix={arXiv},
      primaryClass={cs.AI},
      url={https://arxiv.org/abs/2509.14030}, 
}

@misc{llava,
      title={LLaVA-4D: Embedding SpatioTemporal Prompt into LMMs for 4D Scene Understanding}, 
      author={Hanyu Zhou and Gim Hee Lee},
      year={2025},
      eprint={2505.12253},
      archivePrefix={arXiv},
      primaryClass={cs.CV},
      url={https://arxiv.org/abs/2505.12253}, 
}

@misc{sam2,
      title={SAM 2: Segment Anything in Images and Videos}, 
      author={Nikhila Ravi and Valentin Gabeur and Yuan-Ting Hu and Ronghang Hu and Chaitanya Ryali and Tengyu Ma and Haitham Khedr and Roman Rädle and Chloe Rolland and Laura Gustafson and Eric Mintun and Junting Pan and Kalyan Vasudev Alwala and Nicolas Carion and Chao-Yuan Wu and Ross Girshick and Piotr Dollár and Christoph Feichtenhofer},
      year={2024},
      eprint={2408.00714},
      archivePrefix={arXiv},
      primaryClass={cs.CV},
      url={https://arxiv.org/abs/2408.00714}, 
}

@INPROCEEDINGS{AnnoLens,
  author={Becker, Franziska and Koch, Steffen and Blascheck, Tanja},
  booktitle={IEEE Visualization and Visual Analytics (VIS)}, 
  title={AnnoLens: Exploration and Annotation through Lens-Based Guidance}, 
  year={2025},
  pages={241--245},
  doi={10.1109/VIS60296.2025.00054}}

@misc{sam3,
      title={SAM 3: Segment Anything with Concepts}, 
      author={Nicolas Carion and Laura Gustafson and Yuan-Ting Hu and Shoubhik Debnath and Ronghang Hu and Didac Suris and Chaitanya Ryali and Kalyan Vasudev Alwala and Haitham Khedr and Andrew Huang and Jie Lei and Tengyu Ma and Baishan Guo and Arpit Kalla and Markus Marks and Joseph Greer and Meng Wang and Peize Sun and Roman Rädle and Triantafyllos Afouras and Effrosyni Mavroudi and Katherine Xu and Tsung-Han Wu and Yu Zhou and Liliane Momeni and Rishi Hazra and Shuangrui Ding and Sagar Vaze and Francois Porcher and Feng Li and Siyuan Li and Aishwarya Kamath and Ho Kei Cheng and Piotr Dollár and Nikhila Ravi and Kate Saenko and Pengchuan Zhang and Christoph Feichtenhofer},
      year={2025},
      eprint={2511.16719},
      archivePrefix={arXiv},
      primaryClass={cs.CV},
      url={https://arxiv.org/abs/2511.16719}, 
}

@article{Kelly2024VisionGPT,
  author  = {Chris Kelly and Luhui Hu and Bang Yang and Yu Tian and Deshun Yang and Cindy Yang and Zaoshan Huang and Zihao Li and Jiayin Hu and Yuexian Zou},
  title   = {VisionGPT: Vision-Language Understanding Agent Using Generalized Multimodal Framework},
  journal = {arXiv preprint arXiv:2403.09027},
  year    = {2024},
  url     = {https://arxiv.org/abs/2403.09027}
}

@misc{OpenClaw2026,
  author = {Peter Steinberger},
  title  = {OpenClaw: Open‑Source Autonomous LLM Agent},
  year   = {2026},
  howpublished = {\url{https://openclaw.ai/}},
}

@article{visinfo1,
title = {ClayVolume: A Progressive Refinement Interaction System for Immersive Visualization},
journal = {Visual Informatics},
volume = {9},
number = {1},
pages = {71--83},
year = {2025},
issn = {2468-502X},
doi = {https://doi.org/10.1016/j.visinf.2025.01.003},
url = {https://www.sciencedirect.com/science/article/pii/S2468502X25000038},
author = {Zhenyuan Wang and Qing Zhao and Yue Zhang and Jinhui Zhang and Guihua Shan and Xiao Zhou and Dong Tian},
}

@article{visinfo2,
title = {Interactive Simulation and Visual Analysis of Social Media Event Dynamics with LLM-Based Multi-Agent Modeling},
journal = {Visual Informatics},
volume = {9},
number = {3},
pages = {100260},
year = {2025},
issn = {2468-502X},
doi = {https://doi.org/10.1016/j.visinf.2025.100260},
url = {https://www.sciencedirect.com/science/article/pii/S2468502X25000439},
author = {Zichen Cheng and Ziyue Lin and Yihang Yang and Zhongyu Wei and Siming Chen},
}

@article{visinfo3,
title = {From Perception to Reflection: A Layered Framework for Aesthetic Education in the Digital Design of Ancient Painting},
journal = {Visual Informatics},
volume = {9},
number = {4},
pages = {100290},
year = {2025},
issn = {2468-502X},
doi = {https://doi.org/10.1016/j.visinf.2025.100290},
url = {https://www.sciencedirect.com/science/article/pii/S2468502X25000737},
author = {Xiaojiao Chen and Wenru Qi and Yulian Yang and Xiaosong Wang and Wei Chen}
}

@article{visinfo4,
title = {Reconfiguration of the Brain During Aesthetic Experience on Chinese Calligraphy: Using Brain Complex Networks},
journal = {Visual Informatics},
volume = {6},
number = {1},
pages = {35--46},
year = {2022},
issn = {2468-502X},
doi = {https://doi.org/10.1016/j.visinf.2022.02.002},
url = {https://www.sciencedirect.com/science/article/pii/S2468502X22000109},
author = {Rui Li and Xiaofei Jia and Changle Zhou and Junsong Zhang},
}

@article{visinfo5,
title = {ArtEyer: Enriching GPT-Based Agents with Contextual Data Visualizations for Fine Art Authentication},
journal = {Visual Informatics},
volume = {8},
number = {4},
pages = {48--59},
year = {2024},
issn = {2468-502X},
doi = {https://doi.org/10.1016/j.visinf.2024.11.001},
url = {https://www.sciencedirect.com/science/article/pii/S2468502X24000664},
author = {Tan Tang and Yanhong Wu and Junming Gao and Kejia Ruan and Yanjie Zhang and Shuainan Ye and Yingcai Wu and Xiaojiao Chen},
}

@article{visinfo6,
title = {A Survey of Visual Insight Mining: Connecting Data and Insights via Visualization},
journal = {Visual Informatics},
volume = {9},
number = {4},
pages = {100271},
year = {2025},
issn = {2468-502X},
doi = {https://doi.org/10.1016/j.visinf.2025.100271},
url = {https://www.sciencedirect.com/science/article/pii/S2468502X25000543},
author = {Yijie Lian and Jianing Hao and Wei Zeng and Qiong Luo},
}

@incollection{Brooke1996SUS,
  author       = {John Brooke},
  title        = {SUS: A Quick and Dirty Usability Scale},
  booktitle    = {Usability Evaluation in Industry},
  editor       = {P. W. Jordan and B. Thomas and B. A. Weerdmeester and I. L. McClelland},
  pages        = {189--194},
  year         = {1996},
  publisher    = {Taylor \& Francis},
  isbn         = {9780748404605}
}

@book{Kipling1902JustSo,
  author       = {Rudyard Kipling},
  title        = {Just So Stories},
  year         = {1902},
  publisher    = {Macmillan and Co.},
  isbn         = {978-1509828625},
  note         = {Contains poem ``I Keep Six Honest Serving Men'', source of the 5W1H / Kipling method},
}

@article{Kelley1984WizardOfOz,
  author       = {J. F. Kelley},
  title        = {An Iterative Design Methodology for User‑Friendly Natural‑Language Office Information Applications},
  journal      = {ACM Transactions on Office Information Systems},
  volume       = {2},
  number       = {1},
  pages        = {26--41},
  year         = {1984},
  doi          = {10.1145/357417.357420},
}
\end{document}